\documentstyle[a4,12pt]{article}
\begin{document}
\begin{titlepage} \vspace{0.2in} 

\begin{center} {\LARGE \bf 
Influence of the Particles Creation on the 
Flat and Negative Curved FLRW  Universes  
\\} \vspace*{0.8cm}
{\bf Giovanni Montani
}\\ \vspace*{1cm}
ICRA---International Center for Relativistic Astrophysics\\ 
INFN---Istituto Nazionale di Fisica Nucleare\\ 
Dipartimento di Fisica (G9), Universit\`a  di Roma, ``La Sapienza", 00185 Rome, Italy
\vspace*{1.8cm}

 \vspace*{1cm} \\ 

{\bf   Abstract  \\ } \end{center} \indent
We present a dynamical analysis of the (classical) 
spatially flat and negative 
curved Friedmann-Lame\^{\i}tre-Robertson-Walker (FLRW) 
universes evolving, (by assumption) close to the thermodynamic 
equilibrium, 
in presence of a particles creation 
process, described by means of a realiable  
phenomenological approach, based on the application 
to the comoving volume (i. e. spatial volume of unit comoving 
coordinates) of the theory for open 
thermodynamic systems. 

In particular we show how, since the particles creation phenomenon 
induces a negative pressure term, then 
the choice of a well-grounded {\it ansatz} for the time variation of 
the particles number, 
leads to a deep modification of the very early 
standard FLRW dynamics. 

More precisely for the considered FLRW models, 
we find (in addition to the limiting case of 
their standard behaviours) 
solutions corresponding to an early universe characterized 
respectively by an ``eternal" inflationary-like birth and a 
spatial curvature dominated singularity. 
In both these cases the so-called {\it horizon problem} finds 
a natural solution. 
\end{titlepage}

\section{Introduction} 

Though the ``Standard Cosmological 
Model" \cite{KT90,E} finds its theoretical grounds on a 
purely classical and maximally symmetric solution 
of the Einstein equations, like the 
Friedmann-Lame\^{\i}tre-Robertson-Walker (FLRW) model, nevertheless 
it is commonly believed that quantum effects 
could have played an important dynamical role 
during the evolution of the early universe. 

Among such quantum effects are surely of key relevance 
those related to processes of particles creation; 
indeed, due to the strong time variation singled out by the 
background gravitational field near the cosmological singularity, 
it is naturally expected that quantum fluctuations of the field  
itself, as well as the dynamics of any quantum field living 
on such a nonstationary background, lead to phenomena 
of matter creation, whose existence can in turn modifys the 
early universe behaviour. 
 
A detailed microscopic description of matter creation phenomena 
should be provided by a (fully self-consistent) 
quantum field theory on a curved space-time \cite{BD82}, or, 
when the backreaction of the created matter becomes no longer 
negligible 
(in the sense of strong quantum fluctuations around any average 
effect), by a pure quantum gravity theory in presence of 
``matter" fields. 

A macroscopic phenomenological description of processes of 
matter creation, 
able to make account of the everage effect induced on the 
(classical) dynamics of an isotropic expanding universe,  
was proposed by I. Prigogine et al. (see \cite{P61}-\cite{P89}) 
which is based on the application, to the universe comoving volume 
(i. e. spatial volume of unit comoving coordinates),  
of the standard theory of open thermodynamic systems having 
a conserved specific entropy (entropy per particle); 
such a description was then generalized in 
\cite{C92} by allowing the specific entropy be no longer constant 
and putting its formulation in a completely covariant form.  

If the matter present in the universe is described by the 
energy-momentum tensor of a perfect fluid, then, in such 
an approach, the pressure of the fluid, apart from the usual 
thermostatic component, contains an additional term, directly 
related to the time variation of the particles number within the 
comoving volume; furthermore the total entropy 
within such a comoving volume, 
of the universe, directly proportional to 
the particles number, 
is clearly no longer constant (like in the Standard Cosmological 
Model), since it increases during 
the processes of matter creation 
(with interesting implications about the explanation 
of the actual universe entropy). 

Of course, the above mentioned macroscopic phenomenological 
approach is not completely self-consistent since it can not 
clearly determine the correct 
expression for the rate of particles creation; 
therefore such an information is an open degree of freedom 
of the scheme and it should be provided by 
the formulation of an appropriate {\it ansatz } 
for the considered dynamical model. 

In \cite{D97} the influence of a particles creation term on a 
FLRW universe, having a flat 
three-dimensional space, was studied 
in presence of bulk viscosity, by showing how the assumption of 
a reasonable {\it ansatz} (for the rate of particles creation) 
leads to the appearance of nonsingular cosmologies in correspondence 
to some particular choice of the free parameters of the model. 

Our analysis generalize the {\it ansatz} presented in \cite{D97} 
to make it even suitable for the description of a 
spatially negative curved FLRW model and the resulting early 
dynamics 
(analysed close to the thermodynamic equilibrium and 
in absence of bulk viscosity, which, in agreement to 
what below discussed in section $2$ should be appropriately treated 
in this framework only as a small perturbative term) 
appears to be, in both the cases, as deeply modified 
by the processes of matter creation. Indeed in the case of 
a flat FLRW universe we get a nonsingular cosmology in 
correspondence to an open set of the allowed parameters. 

More precisely, by choosing an appropriate time variable 
(the logaritm of the comoving volume), we 
reduce the dynamical problem to an evolutive equation for the only 
energy density filling the universe, whose analytical solution 
provides a functional form of the time variable which is bounded or 
unbounded depending on the sign of the integration constant. 
In the physical case of a strictly positive integration constant, 
in correspondence to any choice of the allowed parameters, 
the evolution of the universe asymptotically approaches 
for sufficiently late times the FLRW one, while its early stages 
are characterized, respectively in the flat and negative 
curved cases, by an ``eternal'' inflationary-like behaviour and 
a spatial curvature dominated singularity. 

The most important aim of this paper is to show how, 
at least in the adopted phenomenological framework, 
the matter creation 
could have been deeply influenced the very early phases of 
the universe life, up to the point of providing an explanation for 
some fundamental puzzles of the modern cosmology, like the 
{\it cosmological singularity} and the
{\it horizon problem}. 

It is important to observe how, though the precise nature of the phenomena 
of particles creation relays on the detailed microphysics working in the 
early universe, nevertheless the notion of heat Big-Bang 
(indeed we will get below the universe has a maximal initial temperature 
to be reasonably fitted by a value $\ge 10^{19}GeV$) implies that the most 
part of the created particles should be initially photons. 

In section $2$ is given a brief review of the theory of open 
thermodynamic systems, whose application to the universe evolution 
provides the theoretical framework for the dynamical analysis 
developed in section $3$; finally 
concluding remarks follows in section $4$. 

\section{Theory of the Open Thermodynamic Systems} 

Here we provide a description of the theory of open 
thermodynamic systems and its application to the universe 
thermodynamics, which leads to the natural phenomenological 
description of matter creation processes adopted in the analysis 
of section $3$ \cite{P61}-\cite{P89} 
(see also \cite{C92,D97}). 

Let us consider a thermodynamical fluid whose 
energy density, thermostatic pressure, volume, temperature, entropy 
and particles number are 
denoted respectively by $\rho$, $p$, $V$, $T$, $S$ and $N$. 
All these thermodynamical parameters should satisfy the 
Gibbs relation: 

\begin{equation} 
d(\rho V) = TdS - pdV + \mu dN  
\label{c2} 
\end{equation} 

where $\mu$ denotes the chemical potential (i. e. the Gibbs 
function per particle) and is defined as: 

\begin{equation} 
\mu = \frac{(\rho + p)V}{N} - \frac{TS}{N} 
\label{d2} 
\end{equation} 

By the above definition of the chemical potential it is easy to 
rewrite the relation (\ref{c2}) in the form: 

\begin{equation} 
d(\rho V) = NTd\sigma - pdV + \frac{H}{N}dN  
\label{e2} 
\end{equation} 

where $\sigma$ and $H$ denote respectively the specific 
entropy ($\sigma = S/N$) and the total enthalpy 
($H = (\rho + p)V$). 

Now, by the first law of thermodynamics, we get the 
fundamental relation (total energy conservation): 

\begin{equation} 
d(\rho V) = \delta Q - PdV  
\label{f21} 
\end{equation} 

where $\delta Q$ indicates the heat exchanged and $P$  the total 
thermodynamical pressure. 

By the comparison of the two equations (\ref{e2}) and 
(\ref{f21}) take place the following natural identifications: 

\begin{equation} 
\delta Q = NTd\sigma \, \quad P = p + \Pi 
\label{f2} 
\end{equation} 

where $\Pi$ denotes an additional pressure term due to the 
matter creation: 

\begin{equation} 
\Pi = -\frac{H}{N}\frac{dN}{dV} = -(\rho + p)\frac{dlnN}{dlnV} 
\label{g2} 
\end{equation} 
 
When applying this theory to the case of an 
expanding isotropic universe we should require that all the 
thermodynamic functions depend only on time and the volume $V$ 
be the comoving volume corresponding (by convention) 
to an unit coordinate volume. 

Finally, by requiring that the universe 
be an adiabatic system, we get 
the following fundamental relation: 

\begin{equation} 
\delta Q = 0 \Rightarrow NTd\sigma = 0 
\label{h2} 
\end{equation} 

Now, if we treate the universe as a closed system having a fixed 
number of particles per comoving volume 
($N = const.$), then the 
above adiabatic condition (\ref{h2} leads to the entropy 
conservation $dS = 0$ (like in the Standard Cosmological Model); 
instead when the universe behaves 
as an open thermodynamic system and it is allowed 
the particles number per comoving volume changes during its 
evolution, the same adiabatic relation provides the following 
fundamental expression for the entropy variation:    

\begin{equation} 
TdS = \frac{TS}{N}dN 
\label{i2} 
\end{equation} 

To put the above thermodynamical scheme in a covariant form 
(appropriate to our purposes), it is enough to redefine the 
energy-momentum tensor of a perfect fluid as:

\begin{equation} 
T_{ik} = (\rho + P)u_iu_k + Pg_{ik} \, \quad i,k = 0,1,2,3 
\label{l2} 
\end{equation} 

where $u_i$ and $g_{ik}$ denote respectively the four-velocity 
of the fluid and the space-time metric. 

To conclude, it is important to abserve that, since for 
$dlnN/dlnV > (\rho + 3p)/(3\rho + 3p)$ we have 
$\rho + 3P < 0$, then, 
in agreement to \cite{HP}, 
the particles creation term can provide 
a sufficiently negative pressure to remove the cosmological 
singularity. 

Let us now face the question concerning the possibility of 
including in the above scheme viscosity effects. 

Indeed it is commonly believed that phenomena of matter creation 
should be associated with the presence of 
a certain degree of bulk viscosity, as assumed 
in \cite{D97}; however, as discussed in \cite{V}, 
when referred to rapidly space-time varying phenomena 
(like the ones concerning the very early cosmology) 
the standard theory of viscosity 
should be amended for accounting of the finite velocity by which 
any  thermodynamical effect propagates. 

Such an aim can be reached, in a phenomenological approach, 
by introducing suitable relaxation times (typically of the order of 
the mean collision time) which make account for the characteristic 
time of the system evolution. When such characteristic time is 
sufficiently large, then the standard theory of viscous fluids can 
provide a satisfactory description of the considered thermodynamic 
system; but if such time becomes short enough, the system resembles 
more a viscoelastic solid and the causal effects should be taken 
into account. 
 
More precisely, when referred to the only presence of bulk 
viscosity, the theory of viscoelastic matter, derived in \cite{V}, 
can be (dynamically) summarized by the following 
energy-momentum tensor:  

\begin{equation} 
T_{ik} = (\rho + P + \omega )u_iu_k + (P + \omega )
g_{ik} \, \quad i,k = 0,1,2,3 
\label{m2} 
\end{equation} 

where $\omega$ denotes the dilatational-stress coefficient and satisfys 
the equation: 

\begin{equation} 
\omega + \tau _0\dot{\omega} = 
\xi u^i_{;i} \, \quad 
\dot{(\quad )} \equiv (\quad )_{;i}u^i 
\label{n2} 
\end{equation} 

with $;$ indicating the covariant derivative and 
$\tau _0$ and $\xi$ respectively the relaxation time and the 
dilatation coefficent; in the limit $\tau _0\rightarrow \infty$ 
the ratio $\xi /\tau _0$ goes over the bulk modulus. 

Since such a theory is only appropriate to describe 
quasi-stationary systems, which evolve close to the thermodynamical 
equilibrium, then $\omega$ should fulfills the inequality \cite{B}: 
 
\begin{equation} 
\omega ^2\frac{\tau _0}{\xi} \ll \rho 
\label{o2} 
\end{equation} 

Now the first law of the thermodynamics 
(which is provided by the conservation law for the 
energy-momentum tensor (\ref{m2}) and  
in the below analysis plays a key dynamical role) reads: 

\begin{equation} 
d\rho = - (\rho + P + \omega) dlnV 
\label{p2} 
\end{equation} 

Since $\omega$, $\tau _0$ and $\xi$ are expected to be function of 
the only energy density $\rho$ 
(below, by choosing a suitable {\it ansatz} for the particles creation term, 
we get that even $P$ is a function of $\rho$), then 
a good (dimensional) assumption about the form the ratio 
$\xi /\tau _0$ should take, is given by 
\footnote{In \cite{B} it was assumed with satisfactory causal 
results the form $\xi /\tau _0 \sim \rho$.}: 

\begin{equation} 
\frac{\xi }{\tau _0} = \rho \left( 
\frac{\rho }{\rho ^*}\right) ^k \, \quad \rho ^* = const. > 0 
\quad k = const. \ge 0 
\label{q2} 
\end{equation} 

Hence equation (\ref{o2}) rewrites: 

\begin{equation} 
\omega \ll \rho \left( 
\frac{\rho }{\rho ^*}\right) ^{\frac{k}{2}} 
\label{r2} 
\end{equation} 

By the dynamical analysis developed in the next section we 
will show that, in the case of physical interest, during the all 
universe evolution, the energy density satisfys the restriction 
$\rho \le \bar{\rho }\quad \bar{\rho } = const. > 0$. 

Now in the cases 
$\bar{\rho } \ll \rho ^*$ and $\bar{\rho } \sim \rho ^*$, by 
(\ref{r2}), we see that in equation (\ref{p2}) the contribution due 
to  $\omega$ is to be regarded as a small perturbation. 
Instead in the case 
$\bar{\rho } \gg \rho ^*$, the above claim is no longer always true. 
However the former cases appear preferrable from a physical 
point of view, since, 
in the limit of very small values of the viscosity coefficient 
$\xi$, we can yet speak of a real (weakly) viscous fluid 
(i. e. are yet allowed large values of the relaxation time $\tau _0$). 
But the most important physical reason which leads us to believe the 
viscosity effects be negligible in the earliest phases of the universe 
evolution relays on the observation presented at the end of section $1$ 
about the requirement of an early  dominant radiation component; 
indeed termalized photons are not expected to resemble a viscous fluid,  
like instead could take place for a later hadronic component. 

According to the above general statement, 
in our approach we will neglect the viscosity 
effects which, providing only a small dynamical perturbation, 
can not drastically modify the results below obtained 
\footnote{However, on this level, 
we can not exclude at all that, under certain 
conditions, the viscosity effects could play a relevant dynamical role 
during the very early stages of the universe evolution.}. 

\section{Dynamical Model} 

As well known, in a synchronous reference, 
the geometrical structure of the FLRW universe is 
summarized by a line element of the form \cite{KT90}: 

\begin{equation} 
ds^2 = - c^2dt^2 + 
R^2(t)dl^2 
\label{a} 
\end{equation} 

where $c$ denotes the speed of the light, 
$R(t)$ ($t_i \le t \le \infty \, \quad t_i \ge  - \infty$) 
the cosmic scale factor 
(which has the dimension of a lenght) 
and $dl^2$ the line element of 
a constant curvature three-dimensional space, i.e.:

\begin{equation} 
dl^2 = \frac{dr^2}{1 - kr^2} + 
r^2(d\theta ^2 + sin^2\theta d\phi ^2) 
\label{b} 
\end{equation} 

Here by $k$ we indicate the curvature signature 
($k = - 1, 0, + 1$). 

The evolution of the FLRW universe is governed 
by the following 
two Friedmann and ``continuity'' equations: 

\begin{equation} 
H(t)^2\equiv \left( \frac{1}{R}\frac{dR}{dt}\right) ^2 = 
\frac{c^2\chi}{3}\rho - \frac{c^2k}{R^2} 
\label{a2} 
\end{equation} 

\begin{equation} 
d(\rho R^3) = - Pd(R^3) 
\label{b2} 
\end{equation} 

Above $\chi = 8\pi G/c^4$ (being $G$ the Newton constant). 

Now we introduce the new time variable $\eta (t)$, 
defined by: 

\begin{equation} 
d\eta = 3H(t)dt 
\Rightarrow R(\eta ) = R_0e^{\frac{1}{3}(\eta - \eta _0)} 
\, \quad -\infty < \eta < \infty 
\label{d} 
\end{equation} 

where $R_0$ and $\eta _0$ denote two assigned 
constants ($R_0 > 0$). 
By other words we adopt as time variable, the logaritm 
of the comoving volume to describe the universe expansion 
(the above position $d\eta = 3H(t)dt$ 
clearly provides an appropriate time variable only in 
the case of an expanding 
universe, i. e. $H > 0$). 

In terms of (\ref{d}) the line element (\ref{a}) reads: 

\begin{equation} 
ds^2 = - \frac{c^2d\eta ^2}{9H^2(\eta )} +  
{R_0}^2e^{\frac{2}{3}(\eta - \eta _0)}dl^2 
\label{e} 
\end{equation} 

Starting by the above line element and in agreement with 
the thermodynamical scheme presented in the previous section, 
the dynamical equations for the FLRW 
model, in presence of particles creation, take the form 
\cite{KT90,C92}: 

\begin{equation} 
H^2(\eta ) = \frac{c^2\chi }{3} 
\rho (\eta ) - 
\frac{c^2k}{{R_0}^2}
e^{-\frac{2}{3}(\eta - \eta _0)} 
\label{f} 
\end{equation} 

\begin{equation} 
\frac{d\rho }{d\eta } = -(\rho + P) 
\label{g} 
\end{equation} 

where $\rho$ denotes the energy density associated with the matter 
filling the universe, and $P$ the corresponding total 
(isotropic) pressure, constituted by the following two contributions: 

\begin{equation} 
P = p + \Pi \, \quad \Pi \equiv 
- (\rho + p)\frac{dlnN}{dln(R^3)} 
= -(\rho + p)\frac{dlnN}{d\eta } 
\label{h} 
\end{equation} 

Now, to the existence of such dissipative processes is 
associated an entropy production determined, 
in agreement with (\ref{i2}), 
by the relation: 

\begin{equation} 
T\frac{dS}{d\eta } = \frac{TS}{N}\frac{dN}{d\eta } 
\label{i} 
\end{equation} 

Since, according to the second law of the thermodynamics, the entropy never 
decreases ($dS\ge 0$), from (\ref{i}) we immediately get the 
conditions: 

\begin{equation} 
dN \ge 0 \Rightarrow dlnN \ge 0 
\label{l} 
\end{equation} 

In terms of (\ref{h}) equation (\ref{g}) can be rewritten as follows: 

\begin{equation} 
\frac{d\rho }{d\eta } = (\rho + p)\left( 
\frac{dlnN}{d\eta } - 1\right) 
\label{m} 
\end{equation} 

By introducing the particles density 
$n = \frac{N}{{R_0}^3}e^{-(\eta - \eta _0)}$ the above equation 
takes the more compact form: 

\begin{equation} 
\frac{d\rho }{d\eta } = \frac{\rho + p}{n}\frac{dn}{d\eta } 
\label{n} 
\end{equation} 

The above thermodynamical scheme is completed by the 
following evolutive equation 
concerning the universe temperature \cite{C92}: 

\begin{equation} 
\frac{1}{T}\frac{dT}{d\eta } = 
\left( \frac{\partial p}{\partial \rho }\right) _n
\frac{1}{n}\frac{dn}{d\eta } 
\label{o} 
\end{equation} 

We conclude the formulation of the considered cosmological problem 
by observing that the relation between the synchronous time and 
the adopted time variable $\eta$ is given by: 

\begin{equation} 
t(\eta ) = t_0 + \int_{\eta _0}^{\eta }
\frac{d\eta ^{\prime }}{
3H(\eta ^{\prime }) } \, \quad 
t_0 \equiv t(\eta _0) 
\label{pp} 
\end{equation} 

as well as, in terms of our variables, 
the physical horizon reads: 

\begin{equation} 
d_{H}(\eta ) = ce^{\frac{1}{3}\eta } 
\int_{-\infty }^{\eta }\frac{d\eta ^{\prime }}{
3e^{\frac{1}{3}\eta ^{\prime }}H(\eta ^{\prime })} 
\label{q} 
\end{equation} 
 
The advantage in using the adopted time 
variable $\eta$ consists of the decoupled 
character acquired by equation 
(\ref{f}) (i. e. since now $R(\eta )$ is a known function, 
then equations (\ref{f}) amd (\ref{m}) decouple from each other 
\cite{V}), so reducing the 
considered cosmological problem to solve the 
only evolutive 
equation (\ref{m}). 

However to solve the equation (\ref{m}) it is necessary 
to assign the equation of state 
relating the pressure $p$ to the energy density $\rho $, but overall to 
make an appropriate {\it ansatz} for the particles creation term 
$dlnN/d\eta$. 

In what follows we shall assume an equation of state of the 
form \cite{F}: 

\begin{equation} 
p = (\gamma - 1)\rho \, \quad 1 \le \gamma \le 2 
\label{r} 
\end{equation}  

where $\gamma$ denotes the adiabatic index 
(in particular $\gamma $ takes the values $4/3$ for the ultrarelativistic 
matter and $1$ for  a dust). 

In order to individualize an appropriate {\it ansatz} for 
$dlnN/d\eta$, we preliminary observe that, by using (\ref{r}), then 
equation (\ref{n}) immediately gives: 

\begin{equation} 
\rho = \rho _0 \left( \frac{n}{n_0}\right) ^{\gamma } 
\Rightarrow n = n_0\left( \frac{\rho }{\rho _0}
\right) ^{\frac{1}{\gamma }} 
\label{s} 
\end{equation} 

where $\rho_0$ and $n_0$ denote two positive constants 
(namely, $\rho _0 = \rho (\eta _0)$. $n_0 = n(\eta _0)$). 

It is easy to realize that equation (\ref{s}) leads to the following 
implications: 

\begin{equation} 
\rho \equiv 0 \Rightarrow n\equiv 0 \Rightarrow 
\frac{dlnN}{d\eta } = 0 
\label{t} 
\end{equation} 

\begin{equation} 
\rho = const. \ne 0 \Rightarrow n = const. \ne 0 \Rightarrow 
\frac{dlnN}{d\eta } = 1 
\label{u} 
\end{equation} 

Since in the case of a FLRW model with negative curvature ($k = - 1$) 
the condition $\rho \equiv 0$ is yet compatible 
with the existence of a 
real space-time (characterized by $H = c 
exp\{ - (\eta - \eta_0)/3\} /R_0$),  
we are naturally lead to the following {\it ansatz }: 

\begin{equation} 
\frac{dlnN}{d\eta } = \alpha \left( H^2 + \frac{c^2k}{R_0^2}
e^{- \frac{2}{3}(\eta - \eta _0)}
\right)^{\beta } = \left( \frac{\rho }{\bar{\rho }}\right) ^{\beta } 
\label{v} 
\end{equation} 

where $\alpha$ and $\beta$ denote two positive constants and we set: 

\begin{equation} 
\bar{\rho } \equiv \frac{3}{c^2\chi \alpha ^{\frac{1}{\beta }}} 
 \label{w} 
\end{equation} 

This expression for $dlnN/d\eta$ clearly satisfys 
(\ref{t}) and, in view 
of (\ref{u}), if $\rho = \rho _c = const.$, we should have: 

\begin{equation} 
\rho _c = \bar{\rho } 
\label{x} 
\end{equation} 

By using (\ref{v}) and (\ref{r}), 
equation (\ref{m}) can be rewritten 
in the form: 

\begin{equation} 
\frac{d\rho }{d\eta } = \gamma \rho \left[ 
\left( \frac{\rho }{\bar{\rho }}\right) ^{\beta } - 1\right] 
\label{y} 
\end{equation} 

As it is easy to verify the solution of the above equation reads: 

\begin{equation} 
\rho (\eta ) = \frac{\bar{\rho }}{\left( 1 + a\bar{\rho }^{\beta } 
exp\{ \gamma \beta (\eta - \eta _0)\} \right) ^{\frac{1}{\beta }}} 
\label{z} 
\end{equation} 

or, equivalently, in terms of $R$ as: 

\begin{equation} 
\rho (R) = \frac{\bar{\rho }}{\left[ 1 + a\bar{\rho }^{\beta } 
\left( \frac{R}{R_0} \right) ^{3\beta \gamma } 
\right] ^{\frac{1}{\beta }}} 
\label{a3} 
\end{equation} 

where $a$ denotes an arbitrary integration constant. 

By assigning the initial condition $\rho (\eta _0) = \rho _0$, we 
immediately obtain: 

\begin{equation} 
\rho _0 = \frac{\bar{\rho }}{\left( 1 + 
a\bar{\rho } ^{\beta } 
\right) ^{\frac{1}{\beta }}} \Rightarrow 
a = \left( 
\frac{1}{\rho _0 ^{\beta }} - 
\frac{1}{\bar{\rho} ^{\beta }} \right) 
\label{a1} 
\end{equation} 
 
From equations (\ref{s}), (\ref{z}) and (\ref{a1}) we get the 
following expressions for 
the particles density as well as the particles number: 

\begin{equation} 
n = n_0 \left( \frac{\bar{1 + a\bar{\rho} } ^{\beta }}
{1 + a\bar{\rho } ^{\beta } 
exp\{ \gamma \beta (\eta - \eta _0)\} 
} \right) ^{\frac{1}{\gamma \beta }} 
\label{b1} 
\end{equation} 

\begin{equation} 
N = n_0 \left( \frac{(1 + a
\bar{\rho } ^{\beta }){R_0}^{3\gamma \beta } 
exp\{ \gamma \beta  (\eta - \eta_0) \} } 
{1 + a\bar{\rho } ^{\beta } exp\{ \gamma \beta (\eta - \eta _0)\} 
} \right) ^{\frac{1}{\gamma \beta }} \, \quad N_0 = n_0R_0^3 
\label{c1} 
\end{equation} 

Finally, by (\ref{f}), we get the following key expression for 
the expansion rate of the universe: 

\begin{equation} 
H(\eta ) = \sqrt{ 
\frac{c^2\chi }{3} \frac{\bar{\rho }}{\left( 1 + 
a\bar{\rho }^{\beta } 
exp\{ \gamma \beta (\eta - \eta _0)\} \right) ^{\frac{1}{\beta }}} - 
\frac{c^2k}{{R_0}^2}
e^{-\frac{2}{3}(\eta - \eta _0)} } 
\label{d1} 
\end{equation} 

The expression (\ref{z}) shows how the energy density of the 
universe be a nondiverging function of $\eta$ in correspondence 
to the choice $a\ge 0$ ($\rho_0 \le \bar{\rho}$), 
while for $a < 0$ ($\rho_0 > \bar{\rho}$) 
it diverges when $\eta$ 
takes the value 
$\eta _f = \eta _0 + [ln(-1/a\bar{\rho }^{\beta })]/\gamma \beta $. 
However, for any value of the constant $a$, in the limit 
$\eta \rightarrow -\infty$, we get $\rho \rightarrow \bar{\rho }$. 

As preliminary step, it is important to stress that the two cases 
$a = 0$ (Bondi-Gold steady state universe) 
and $a < 0$, by describing solutions in which 
the energy density respectively remains constant or increases 
monotonically (diverging for $\eta = \eta _f$), do not provide a 
suitable representation of our actual universe and therefore 
will be excluded from the following analysis. 

Then, it is easy to realize that in the limiting case when 
the processes of matter creation are absent 
($\alpha = 0, i. e. \bar{\rho} \rightarrow \infty$), we 
immediately reobtain, by (\ref{d1}), (\ref{pp}) and (\ref{a3}), 
the standard FLRW dynamics. Furthermore our solution 
($a > 0$) goes 
over the FLRW dynamics in the limit $\eta \rightarrow \infty$, 
corresponding to the behaviour: 

\begin{equation} 
\rho \propto \frac{1}{R^{3\gamma }} 
\label{b3} 
\end{equation} 

Indeed the key difference between the Bondi-Gold steady state 
model and the our one, consists just of such asymptotic FLRW 
dynamics, 
which allows to overcome the two fundamental cosmological 
inconsistencies of a steady universe, concerning the existence 
of a microwaves background radiation and the unobserved level 
of particles creation, as required to keep constant 
the present energy density of the universe. 

More precisely, it is easy to realize that, 
by a suitable finetuning of the free parameters 
$\bar{\rho }$, $\beta $ and $a$, it is possible to fit 
all the fundamental (predictive) 
features of the Standard Cosmological Model, 
including the nucleosynthesis process. 

Let us now analyse separately the two cases 
$k = 0$ and $k = - 1$ corresponding to the choice $a > 0$. 

\vspace{0.5cm} 

i) {\it Flat FLRW model} 

In the case of a flat FLRW model ($k = 0$) the function $H(\eta )$ 
behaves as the square root of $\rho (\eta )$ and, as it is easy to 
realize, it 
vanishes for $\eta \rightarrow \infty$ while for $\eta \rightarrow - 
\infty$ approaches the value $c\sqrt{\chi \bar{\rho }/3}$. 
Thus, in the limit $\eta \rightarrow - \infty$, 
we get the following expressions: 

\begin{equation} 
\eta - \eta _0 \sim c\sqrt{3\chi \bar{\rho }} (t - t_0) 
\Rightarrow R(t) \sim R_0exp \left\{ c\sqrt{\frac{\chi \bar{\rho }}{3}} 
(t - t_0) \right\} \sim 0 
\label{d1n} 
\end{equation}

\begin{equation} 
\rho \sim \bar{\rho} \Rightarrow n \sim \bar{n} 
\equiv n_0 \left( 
\frac{\bar{\rho }}{\rho _0}\right) ^{\frac{1}{\gamma }} 
\, \quad N\sim 0 
\label{e1} 
\end{equation} 

In addition it comes out easily from (\ref{pp}) and (\ref{q}) that the 
age of the universe and the physical horizon have a 
diverging character. 

Thus we see how the universe emerges from $\eta = - \infty$ by an 
inflationary-like behaviour characterized by a 
(nearly) constant value of the energy 
and particles number densities, to be interpreted as the following 
asymptotic limits: 
 
\begin{equation} 
\bar{\rho } = \lim_{\eta \rightarrow - \infty } \frac{U}{V} 
\, \quad 
\bar{n} = \lim_{\eta \rightarrow - \infty } \frac{N}{V} 
\label{f1} 
\end{equation} 

being $V$ the considered comoving volume ($V = R^3$) and 
$U$ the total internal energy in it 
contained ($U \equiv \rho R^3$). 

By other words, though $U$ 
and $N$ vanish as $t\rightarrow - \infty $, 
nevertheless their ratio to the 
vanishing comoving volume $R^3$ approaches constant values.

\vspace{0.5cm} 

ii) {\it Negative curved FLRW model} 

In the case of a negative curved FLRW model ($k = - 1$) we see from 
(\ref{d1}) how the function $H(\eta )$ vanishes for $\eta \rightarrow 
\infty$ and diverges 
for $\eta \rightarrow - \infty$. 
As it is easy to verify, for $\eta \rightarrow - \infty$ 
take place the 
following relations: 

\begin{equation} 
\eta - \eta _0 \sim 3ln\left( \frac{R_0 + c(t - t_0)}{R_0}\right) 
\Rightarrow R(t) \sim R_0 + c(t - t_0) \sim 0 
\label{g1} 
\end{equation}

\begin{equation} 
\rho \sim \bar{\rho} \Rightarrow n \sim \bar{n} \equiv n_0 \left( 
\frac{\bar{\rho }}{\rho _0}\right) ^{\frac{1}{\gamma }} \, \quad 
N\sim 0 
\label{i1} 
\end{equation} 

Above, the values $\bar{\rho }$ and $\bar{n}$ 
should be interpreted in terms of the limits (\ref{f1}). 

By (\ref{q}) it is again easy to realize that the physical horizon 
diverges even in the case of a negative curved FLRW universe, 
while by (\ref{pp}) the age of the universe is in this case finite 
$\sim R_0/c + t - t_0$.  

Since, asymptotically, the energy density approaches the constant value 
$\bar{\rho }$, then 
we see how the very early dynamics results to be dominated by the 
spatial curvature (i. e. $H \propto 1/R$). 

\vspace{0.5cm} 
 
In both the above cases ($k = 0$ and $k = - 1$), since, for 
$\eta \rightarrow \infty$ ($t \rightarrow \infty$) 
the energy density 
vanishes and the appropriate equation of state corresponds to 
$\gamma = 1 $, then. in this limit, we get 
$N \sim n_0{R_0}^3a^{- 1/\beta }/\rho _0 = 
N_0a^{- 1/\beta }/\rho _0$. 

At the end we observe, how by (\ref{i}), (\ref{o}) 
and (\ref{r}), the entropy and the temperature of the universe 
can be expressed respectively in terms of the particles 
number and number density as follows: 

\begin{equation} 
S = S_0\frac{N}{N_0} 
\, \quad 
T = T_0\left( \frac{n}{n_0}\right) ^{\gamma - 1} 
\label{n1} 
\end{equation}

By the above expressions for 
$N(\eta )$ and $n(\eta )$, it is easy to construct the explicit 
functional form for $S(\eta )$ and $T(\eta )$. 
In particular 
it is worth noting how in the two cases $k = 0$ and $k = - 1$ 
the entropy reaches for $t\rightarrow \infty$ the 
(maximum) value $S \sim S_0a^{- 1/\beta }/\rho _0$, 
while the (maximum) initial value taken by the temperature is 
given by: 

\begin{equation} 
T_{max} = T_0\left( 
\frac{\bar{\rho }}{\rho _0}\right) ^{1 - \frac{1}{\gamma }} 
\label{o1} 
\end{equation} 

In particular for an initial radiation dominated universe 
($\gamma = 4/3$) we get the maximum value 
$T^r_{max} = T_0 (\bar{\rho }/\rho_0)^{1/4}$. 

Of course the values of the above maximal temperature 
can be suitably fixed to avoid any contraddiction with the 
observed universe. 

\section{Concluding Remarks} 

Finally we enphasize how 
the analysis of a positive curved FLRW universe 
($k = 1$) can not be developed in terms of our approach since 
the time variable $\eta$ would be hill-defined in the 
``turning point'' of the universe ($H = 0$) 
(as well as not applicable to the collapsing picture $H < 0$), 
naturally 
expected in the evolution of such a model. 
Furthermore, as it is easy to realize, for 
negative values of $H$ the adopted 
{\it ansatz} would lead 
to a negative synchronous time rate of particles and entropy 
production, contraddicting the second law of thermodynamics; 
for these reasons, the analysis of this case requires  further 
investigation. 

As the results of our analysis show, the presence of a 
particles creation term deeply influences the universe dynamics, 
by allowing the energy density and the temperature have 
never  diverging character. 
More precisely we get, as 
above outlined, a drastic departure from the 
``Standard Cosmology" only in the (very) early stages of the 
evolution, while 
for sufficiently later times (depending on the adopted choice of 
the free parameters), 
the obtained dynamics approaches the 
usual FLRW one. 

It is worth enphasizing how, 
as a direct consequences of this departure, 
the obtained solutions for the two cases 
$k = 0$ and $k = - 1$ both provide, 
by the diverging character of the 
physical horizon, 
a solution to the so-called {\it horizon 
problem}. 

We conclude by observing how any physical motivation for 
the adopted {\it ansatz}
 is rather difficult to be provided 
since it should be referred to an average procedure 
over all quantum effects being active in the early universe. 
In particular the impossibility to make precise account for the 
quantum gravity effect limits strongly the physical 
derivation of the particles creation rate 
(indeed, in the framework of our phenomenological 
approach, the matter is created by the gravitational field itself 
by its time variation and the main contribution to this 
average effect can be expected to come from the quantum fluctuations 
of the cosmological background). 

However, in spite of this great difficulties,  the 
relevant influence the matter creation phenomena 
could have exerted on the very early universe dynamics, 
we believe, makes of interest any effort in this direction. 

\vspace{1cm} 

Daniele Oriti and one of the two referees are thanked for their valuable 
advices on this matter. 

This paper was supported, in part by ICRA and mainly by INFN. 

{99} 

\end{document}